\documentclass[a4paper]{jpconf}
\usepackage{graphicx,amsmath,alltt,axodraw2}

\DeclareSymbolFont{ttoperators}{OT1}{cmtt}{m}{n}
\newcommand\xCode[1]{{%
  \mathcode`\"="0\the\symttoperators22%
  \mathchardef\$="4\the\symttoperators24%
  \mathcode`\(="4\the\symttoperators28%
  \mathcode`\)="5\the\symttoperators29%
  \mathcode`\/="0\the\symttoperators2F%
  \mathcode`\[="4\the\symttoperators5B%
  \mathcode`\]="5\the\symttoperators5D%
  \mathchardef\{="4\the\symttoperators7B%
  \mathchardef\}="5\the\symttoperators7D%
  \ensuremath{\mathtt{#1}}}}
\newcommand\Code[1]{\texttt{#1}}
\newcommand\Var[1]{\textit{#1}}
\newcommand\eg{e.g.\ }
\newcommand\ie{i.e.\ }
\newcommand\rd{\mathrm{d}}

\newcommand{\ket}[1]{\left| #1\right\rangle}
\newcommand{\bra}[1]{\left\langle #1\right|}
\renewcommand\Re{\mathop{\mathrm{Re}}}

\begin{document}

\title{FormCalc 8: Better Algebra and Vectorization}

\author{B.~Chokoufe Nejad$^1$, T.~Hahn$^2$, J.-N.~Lang$^1$, E.~Mirabella$^2$}

\address{%
${}^1$ Universit\"at W\"urzburg,
       Institut f\"ur Theoretische Physik und Astrophysik,
       D--97074 W\"urzburg, Germany \\
${}^2$ Max-Planck-Institut f\"ur Physik,
       F\"ohringer Ring 6,
       D--80805 Munich, Germany}

\ead{%
\{Bijan.Chokoufenejad,Jean-Nicolas.Lang\}@physik.uni-wuerzburg.de,
\{hahn,mirabell\}@mpp.mpg.de%
}

\begin{abstract}
We present Version 8 of the Feynman-diagram calculator FormCalc.
New features include in particular significantly improved algebraic
simplification as well as vectorization of the generated code.
The Cuba Library, used in FormCalc, features checkpointing to
disk for all integration algorithms.
\hfill Report MPP-2013-273
\end{abstract}

\section{Introduction}

The Mathematica package FormCalc \cite{FormCalc} simplifies Feynman 
diagrams generated with FeynArts \cite{FeynArts} up to one-loop order.  
It provides the analytical results and can generate Fortran code for the 
numerical evaluation of the squared matrix element.  Cuba is a library 
for multidimensional numerical integration which is included in FormCalc 
but can also be used independently.

This note presents the following features new in FormCalc 8 and Cuba 
3.2:
\begin{itemize}
\item Significant improvement of the algebraic simplification with
  FORM 4 features.

\item Vectorization of the helicity loop.

\item Automated C-code generation.

\item Optimizations for unitarity methods.

\item Checkpointing for all Cuba algorithms.
\end{itemize}

\section{Improvements in the Algebraic Simplification}

The algebraic simplification of Feynman amplitudes is split between 
Mathematica and FORM.  In a preprocessing stage, Mathematica translates 
the elements of a FeynArts amplitude into FORM syntax and writes them to 
an input file for FORM.  (Note that none of the FeynArts symbols are 
directly redefined, such that processing does not start automatically.)  
FORM then does the major part of the symbolic simplification.  In a
postprocessing step, the FORM output is read and returned to Mathematica 
by FormCalc's ReadForm MathLink utility.

Many new and useful features were introduced in FORM 4 \cite{FORM4}, 
most notably abbreviationing and factorization.  The FORM part of 
FormCalc 8 has been rewritten to take advantage of these facilities, 
resulting in significantly improved algebraic simplification.

\subsection{Abbreviationing}

Once a partial expression is considered final at a particular point in 
the FORM program it is abbreviated, \ie substituted by a symbol.  This 
not only shortens the active expressions but makes the abbreviated parts 
inert, such that subsequent id-statements do not spend time on matching 
these, thus making the FORM code run faster.

A similar technique has been used since Version 6 \cite{FC6}, where the 
FORM expressions were sent on a round-trip to Mathematica halfway 
through the evaluation for introducing abbreviations.  Since this 
involved quite some transmission overhead, it was performed only once 
during each FORM run.  With abbreviationing built into FORM now, 
abbreviations are introduced whenever possible, thereby obviating the 
extra pass to Mathematica.

Abbreviationing also serves to prevent FORM's automatic expansion of 
expressions, \ie it preserves a (pre)factorized structure, which is 
particularly useful in combination with the new factorization available 
in FORM (see Sect.~\ref{sect:fact} below).

What is more, since Mathematica receives an expression in many small 
pieces rather than one large chunk, more aggressive simplification 
functions can be applied upon return to Mathematica at reasonable 
efficiency.  To this end, FormCalc wraps a zoo of simplification 
functions around various parts of the amplitude.  All of these are 
`transparent' in the sense that they can be replaced by \Code{Identity} 
without affecting the numerical result.  The three most important ones 
are listed below, a complete inventory is given in the FormCalc manual.
\begin{itemize}
\item \Code{FormSub} is applied to subexpressions of an amplitude.
\item \Code{FormDot} is applied to combinations of dot products in an
  amplitude.
\item \Code{FormMat} is applied to the coefficients of matrix elements
  (`\Code{Mat}') in an amplitude.
\end{itemize}

On the technical side, since the abbreviations are also transmitted to 
Mathematica as such (\ie not back-substituted into the expressions), the 
volume of data transferred is significantly reduced and the final 
expression is stored efficiently as multiple instances of a 
subexpression are taken care of by reference count (same as with 
\Code{Share[]}).

At the moment, FormCalc does not use FORM 4's ``\Code{format~O\Var{n}}''
output optimization, as it is not yet clear how to combine it with the 
postprocessing in ReadForm and Mathematica.

\subsection{Factorization}
\label{sect:fact}

FORM's new `full' factorization (over the rationals) makes it possible 
to simplify expressions much better already inside of FORM.  The old 
`simple' factorization (pulling out common symbols from an expression) 
is still used in instances where full factorization is too expensive.

Potentially time-consuming instances of the \Code{factarg} command in 
the FORM code can be suppressed by setting the \Code{CalcFeynAmp} option 
\Code{NoCostly\,$\to$\,True}.  This is occasionally necessary in models
with more complex couplings such as the MSSM.

Plain factorization is not a cure-all for arbitrary expressions, 
however.  For example, while the following expression is not 
factorizable as a whole,
\begin{verbatim}
-2*e2.k5*S35 + 2*e2.k5*T24 + 2*e2.k5*T14 - 2*e2.k5*MT2 +
2*e2.k5*S - 3*e2.k6*S35 - e2.k6*S45 - e2.k6*T25 - e2.k6*T15 + 
4*e2.k6*T24 + 4*e2.k6*T14 - 4*e2.k6*MT2 + 4*e2.k6*S
\end{verbatim}
it easily admits further compactification by collecting with respect to 
the dot products first:
\begin{verbatim}
-2*(MT2 - S + S35 - T14 - T24)*e2.k5 - 
(4*MT2 - 4*S + 3*S35 + S45 - 4*T14 + T15 - 4*T24 + T25)*e2.k6
\end{verbatim}
FormCalc takes typical objects such as dot products into account, of 
course.  Still, for a general expression it is not straightforward to 
find a suitable simplification procedure, which is why it is useful 
to have functions like \Code{FormDot} through which one can apply 
more sophisticated functions such as Mathematica's \Code{Simplify}.

\section{Vectorization of the Helicity Loop}

The assembly of the squared matrix element in FormCalc can be sketched 
as in the following figure, where the helicity loop sits at the center 
of the calculation:
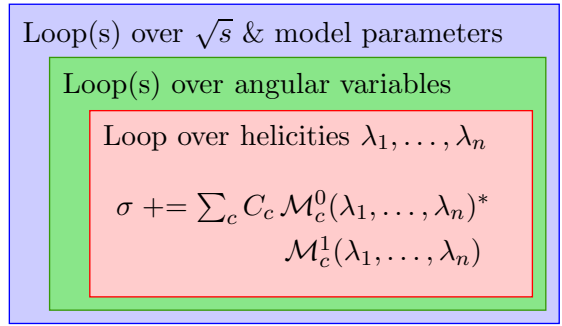
\begin{center}
\begin{picture}(205,120)
\CBox(0,0)(205,120){Blue}{PastelBlue}
\Text(5,115)[tl]{Loop(s) over $\sqrt s$ \& model parameters}
\CBox(15,5)(200,100){OliveGreen}{PastelGreen}
\Text(20,95)[tl]{Loop(s) over angular variables}
\CBox(30,10)(195,80){Red}{PastelRed}
\Text(35,75)[tl]{Loop over helicities $\lambda_1,\dots,\lambda_n$}
\Text(40,50)[tl]{$\sigma \mathrel{{+}{=}}
  \sum_c C_c\,\mathcal{M}^0_c(\lambda_1,\dots,\lambda_n)^*$}
\Text(103,33)[tl]{$\mathcal{M}^1_c(\lambda_1,\dots,\lambda_n)$}
\end{picture}
\end{center}
The helicity loop is not only strategically the most desirable but also 
the most obvious candidate for concurrent execution, as FormCalc does 
not insert explicit helicity states during the algebraic simplification 
\cite{FCopt}.  That is, the amplitude is a numerical function of the 
helicities $\lambda_i$ and not a bunch of (different) functions for 
each helicity combination,
$$
\mathcal{M} = \mathcal{M}(\lambda_1, \lambda_2, \dots)
\neq \{\mathcal{M}_{--\cdots},\ \mathcal{M}_{+-\cdots},\
     \mathcal{M}_{-+\cdots},\ \mathcal{M}_{++\cdots}\}\,.
$$
Such a design is known as Single Instruction Multiple Data (SIMD) in 
computer science since a single code ($\mathcal{M}$) is independently 
run for multiple data ($\lambda_i$), and is conceptually easy to 
parallelize or vectorize.

Parallelization on the CPU's cores using \Code{fork}/\Code{wait} has 
been available from Version 7.5 on \cite{FC75}.  The drawback of this 
method is that it competes for compute cores in particular with Cuba.  
For better efficiency the cores should be assigned to Cuba since it 
computes entire phase-space points in parallel, not just the helicity 
loop.

GPU parallelization was attempted using OpenCL but we found that it was 
not too efficient.  Since the transfer of data between the CPU and the 
GPU is relatively time-consuming, we believe that the distribution of 
the helicity-independent variables from the CPU to the GPU outweighed 
the parallelization gains.

Eventually the best speedup we could achieve was with vectorization.  
With Intel's x86 vector instructions, there is essentially no overhead.

\subsection{Vectorization in C}

Our implementation in C is based on the vector data type extensions 
offered by gcc and Intel's icc.  Unfortunately, these are restricted to 
real algebra, even in C99.

Complex addition is obviously no problem and complex multiplication 
might have been solved through C++'s operator overloading.  Since we 
wanted to stick to C to avoid linking hassles with Fortran object files, 
we adjusted the C-code generation in Mathematica to insert explicit 
macros for the multiplication of complex vectors: \Code{SxH} stands for 
``scalar times helicity vector'' and \Code{HxH} for ``helicity vector 
times helicity vector.''  Helicity vectors are declared with 
\Code{HelType}.  (To avoid confusion with Minkowski four-vectors, we use 
prefixes `\Code{Hel}' or `\Code{H}' to denote helicity vectors in the 
SIMD sense.)

Depending on the hardware features indicated by preprocessor flags these 
macros emit explicit SSE3 or AVX instructions.  For SSE3 the maximum 
vector length is 1 (2 doubles per operation) which may at first not seem 
very useful, but besides performing addition twice as fast there exists 
an efficient complex multiplication routine with 2.5 instructions 
instead of 6.  For AVX (requires i7 `Sandy Bridge' or higher) the 
maximum vector length is 2 (4 doubles per operation).  Again the complex 
multiplication can be formulated fairly efficiently using Intel's vector 
instructions. Overall we found a speedup of 3.7 out of theoretical 4 
with AVX for the helicity loop.

Currently the configure script does not automatically add flags to 
switch on SSE3 or AVX instructions, \eg gcc needs the extra flag 
\Code{-march=corei7-avx} to enable the latter.  This may change in the 
future.  A related question is which default to choose for executables 
that could potentially be run on a cluster of computers with differing 
SIMD capabilities.

\subsection{Vectorization in Fortran}

Vector data types are standard fare in Fortran 90 and so not only 
complex vectors are allowed but one can, in principle, choose arbitrary 
vector lengths.  On the downside, the actual deployment of vector 
instructions is at the discretion of the compiler and may not be chosen 
for vector lengths incommensurable with the hardware.

Even though Fortran 90 is an effective requirement for vectorized 
computation, the code is still generated in fixed format and can be made 
compatible with Fortran 77 through preprocessor definitions, \eg for 
inclusion in legacy packages.

\section{Automated C-Code Generation}

C-code generation has been available from FormCalc 7 on \cite{FC7} but 
now its use is mostly automatic, \ie also drivers and utility files are 
available in C.  In fact, only the declarations needed to be translated 
as the initialization still takes place in Fortran and the C object 
files are simply linked in.  For this to work, the layout of C's structs 
must match Fortran's common blocks, of course.  Private declarations, 
\eg for new models, are not automatically translated, but this is fairly 
straightforward as can be seen by comparing the C and Fortran versions 
of \eg the Standard Model declarations.

To switch from Fortran to C output, the following statement needs to 
precede the output commands (\eg \Code{WriteSquaredME}, 
\Code{WriteRenConst}):
\begin{verbatim}
  SetLanguage["C"]
\end{verbatim}
The output is by default in C99, because of complex numbers, but can 
easily be made to work with C++ by redefining the abstract data type
\Code{ComplexType}.

Even without SIMD vectorization, and perhaps remarkably so for Fortran 
aficionados, C and Fortran versions of the same amplitude show very 
similar performance figures, \ie there is no penalty for using C.

\section{Optimizations for Unitarity Methods}

FormCalc can generate amplitudes for evaluation with the OPP (Ossola, 
Papadopoulos, Pittau \cite{OPP}) unitarity method as implemented in the 
two libraries CutTools \cite{CutTools} and Samurai \cite{Samurai}.  
Instead of introducing tensor coefficients \cite{PaVe}, the whole 
numerator is placed in a subroutine, as in:
\begin{align*}
\varepsilon_1^\mu\varepsilon_2^\nu B_{\mu\nu}(p, m_1^2, m_2^2)
&= (\varepsilon_1\cdot\varepsilon_2) B_{00} +
   (\varepsilon_1\cdot p) (\varepsilon_2\cdot p) B_{11}
&\text{(tensor coeff.)} \\
&= B_{\mathrm{cut}}(2, N, p, m_1^2, m_2^2)\,,
&\text{(OPP)} \\[1ex]
\text{where}\quad
N(q_\mu) &= (\varepsilon_1\cdot q) \: (\varepsilon_2\cdot q)\,.
\end{align*}
The numerator subroutine $N$ will be sampled by the OPP function 
($B_{\mathrm{cut}}$ in this example).  The first argument of
$B_{\mathrm{cut}}$, 2, refers to the maximum power of the integration 
momentum $q$ in $N$.

The OPP procedure indeed generates significantly fewer terms than the 
traditional Passarino--Veltman decomposition, nevertheless a naive 
implementation runs quite a bit slower than its counterpart with tensor 
coefficients.  This section describes our attempts to optimize the OPP 
performance.  We were able to bring the slowdown from originally a 
factor 10 to about a factor 3 for multiplicities such as $2\to 3$ and 
hope to improve matters further.  To be fair, OPP was in the first place 
designed to increase the reach of one-loop calculations to higher-leg 
multiplicities and not so much to speed up the ones with not so many 
legs.

\begin{itemize}
\item
The major part of the slowdown (at least half of that factor 10) comes 
from the fact that the OPP master integrals (the scalar integrals $A_0$, 
$B_0$, $C_0$, $D_0$) are naively computed over and over again.  This is 
because the OPP functions must be evaluated inside the helicity loop 
since the numerator subroutine depends on the helicities.  The scalar 
integrals contain only the denominators, however, and thus could simply 
be moved outside the helicity loop.

In FormCalc, we generate code that foresees a split between the 
computation of the masters and their use in assembling the tensor 
integrals, for example:
\begin{verbatim}
  ComplexType mas145(Mcc)
  ...
  call Cmas(mas145, (C0 args))
  ...
  call Ccut(mas145, num, (C0 args))
\end{verbatim}
The complex array \Code{mas145} stores the master integrals computed by 
\Code{Cmas} (outside the helicity loop) and used by \Code{Ccut} (inside 
the helicity loop).  Unfortunately, so far none of the available OPP 
libraries allows this decomposition even though the two tasks 
`\Code{Cmas}' and `\Code{Ccut}' must be completed internally in some way 
or another already now.  LoopTools alleviates the situation by 
retrieving recurring masters from its cache, though even here the lookup 
time could be eliminated with the above construction.

\item
Some packages address this problem by moving the helicity sum into the
numerator.  This works if only the interference term is sought since 
then the amplitude contains at most one loop integration in each term:
$$
\sum_\lambda 2\Re\mathcal{M}_0^*
  \underbrace{\int\rd^4 q\frac{N}{D\cdots}}_{\sim \mathcal{M}_1} =
\int\rd^4 q\frac{\sum_\lambda 2 \Re\mathcal{M}_0^* N}{D\cdots}
$$
In FormCalc we don't pursue this strategy, firstly because it is not 
applicable if the tree-level contribution is zero (or so small that 
including the loop-squared part becomes necessary) and secondly because 
it is not obvious how this evaluation fits into the present abbreviation 
concept.

\item
Subexpressions of the numerator function (coefficients, summands, etc.) 
independent of $q$ are pulled out and computed once, ahead of invoking 
the OPP function, using FormCalc's abbreviationing machinery 
\cite{FCabbr}.  In particular in BSM theories, these coefficients can be 
lengthy such that pulling them out significantly increases performance.

\item
Our implementation admits mixing Passarino--Veltman decomposition with
OPP in the sense that one chooses an integer $n$ starting from which an
$n$-point function is treated with OPP methods.  For example, \Code{OPP
$\to$ 4} means that $A$, $B$, $C$ functions are treated with
Passarino--Veltman and $D$ and up with OPP.

\item
We optimize OPP calls to reduce sampling effort, \eg by collecting 
denominators, as in:
$$
\frac{N_4}{D_0 D_1 D_2 D_3} + \frac{N_3}{D_0 D_1 D_2}
\to \frac{N_4 + D_3 N_3}{D_0 D_1 D_2 D_3}
$$
Depending on $N$ and rank, joining integrals is not universally better.  
Rather, we tabulated the sampling behavior of Samurai and CutTools such 
that the algorithm can determine the optimal splitting.

\item
The Ninja library implements the $D$-dimensional integrand reduction via 
Laurent expansion \cite{Ninja}, which constructs the tensor integral 
from fewer samples in a numerically more stable way.  Ninja requires a 
slightly modified numerator subroutine which is currently being 
implemented in FormCalc.

\item
The profiler pointed us to a bottleneck in fermion chains: Before, we 
were using elementary operations to build up the fermion chain, which we 
have now merged into a single inlined function call:
\begin{align*}
\bra{u}\sigma_\mu\overline{\sigma}_\nu\sigma_\rho\ket{v}
k_1^\mu k_2^\nu k_3^\rho
&= \bra{u} k_1 \overline{k}_2 k_3\ket{v} \\
\text{(old)} &= \Code{SxS(}u,\,  
     \Code{VxS(}k_1,\,
       \Code{BxS(}k_2,\,
         \Code{VxS(}k_3,\, v
   \Code{))))} \\
\text{(new)} &= \Code{ChainV3(}
   u, k_1, k_2, k_3, v
   \Code{)}\,.
\end{align*}
The elementary operations could not be inlined in Fortran because they 
returned a 2-component spinor, not a scalar value.

\item
The helicity information of an OPP integral's prefactor is taken into 
account in an extra argument.  That is, if a term in the amplitude is 
known to become zero for a particular helicity combination due to its 
prefactor (because of massless external particles, say), the evaluation 
of the loop integral therein is cut short.  For example, the following 
\Code{Dcut} is actually computed only if $\Code{Hel1}\neq 1$:
$$
\Code{Dcut}(3, N, 1 - \Code{Hel1}, \dots)
$$

\end{itemize}

\section{Checkpointing in Cuba}

Cuba's current Version 3.2 allows checkpointing for all routines.  
Checkpointing means writing out the integrator's complete state to disk 
to be able to recover from the last state after a crash.  In a 
long-running calculation this may mean losing one hour instead of one 
day.

Checkpointing is enabled by specifying a name for the state file.  Note 
that, since only Vegas had this functionality in Version 3.0, the 
invocation of the other routines has changed to incorporate the extra 
state file argument.  We always write the state to a new file and remove 
the former state file only when the new one is successfully stored.  
This makes checkpointing failsafe since even a crash during saving is 
recoverable.  If the integration finishes successfully, the state file 
is removed.

The checkpoints have been implemented in the serial regions of the code 
which ensures reliable behavior regardless of parallelization.

Version 3.2 furthermore relaxes several restrictions on the compiler, it 
is now fully C99-compliant and uses no gcc extensions.

\section{Summary}

FormCalc 8 (\Code{http://feynarts.de/formcalc}) has many new and 
improved features, most notably better algebra, a vectorized helicity 
loop, and OPP improvements.
The Cuba library (\Code{http://feynarts.de/cuba}), also included in 
FormCalc, adds checkpointing for all four integration algorithms,
which is useful for resuming interrupted long-running integrations.

\medskip
\raggedright

\end{document}